\documentclass[reprint,aps,prl,twocolumn,superscriptaddress,preprintnumbers,letterpaper,natbib,floatfix,nofootinbib]{revtex4-1}
\usepackage{amssymb}
\usepackage{amsmath}
\usepackage{epsfig}
\usepackage{comment}
\usepackage{color}
\usepackage[hidelinks]{hyperref}
\usepackage{cleveref}
\usepackage{multirow}
\usepackage{capt-of}
\usepackage{siunitx}
\usepackage{graphicx}
\usepackage{slashed}
\usepackage{tabularx}
\usepackage{enumitem}
\usepackage{multirow}
\usepackage{booktabs}
\usepackage{isotope}
\usepackage{xcolor}
\usepackage{bm}
\usepackage{array,mathtools}
\usepackage{fontawesome}

\newcommand{\prn}[1]{ \left(  #1 \right) }

\newcommand{\Neff}{N_\text{eff}}
\newcommand{\al}[1]{\begin{align} #1 \end{align} }
\newcommand{\mysec}[1]{\paragraph*{#1.}\!\!\!\!\!---}
\newcommand{\avg}[1]{\langle #1 \rangle}
\newcommand{\gamp}{{\gamma '}}

\newcommand{\sigv}{\avg{\sigma v}}

\newcommand{\xxtopsps}{\chi \overline{\chi} \to \psi \overline{\psi}}
\newcommand{\smsmtoxx}{\text{SM} \; \text{$\overline{ \rm SM}$} \to \chi \overline{\chi}}
\newcommand{\sige}{\overline{\sigma}_e}

\begin{document}

\title{A Dark Sink Enhances the Direct Detection of Freeze-in Dark Matter}

\author{Prudhvi N.~Bhattiprolu}
\affiliation{Leinweber Center for Theoretical Physics, Department of Physics,\\ University of Michigan, Ann Arbor, MI 48109, USA}
\author{Robert McGehee}
\affiliation{William I. Fine Theoretical Physics Institute, School of Physics and Astronomy,\\ University of Minnesota, Minneapolis, MN 55455, USA}
\affiliation{Leinweber Center for Theoretical Physics, Department of Physics,\\ University of Michigan, Ann Arbor, MI 48109, USA}
\author{Aaron Pierce}
\affiliation{Leinweber Center for Theoretical Physics, Department of Physics,\\ University of Michigan, Ann Arbor, MI 48109, USA}

\begin{abstract}
We describe a simple dark sector structure which, if present, has implications for the direct detection of dark matter (DM): \emph{the Dark Sink}. A Dark Sink transports energy density from the DM into light dark-sector states that do not appreciably contribute to the DM density. As an example, we consider a light, neutral fermion $\psi$ which interacts solely with DM $\chi$ via the exchange of a heavy scalar $\Phi$. We illustrate the impact of a Dark Sink by adding one to a DM freeze-in model in which $\chi$ couples to a light dark photon $\gamp$ which kinetically mixes with the Standard Model (SM) photon. This freeze-in model (absent the sink) is itself a benchmark for ongoing experiments. In some cases, the literature for this benchmark has contained errors; we correct the predictions and provide them as a public code \href{https://github.com/prudhvibhattiprolu/FreezeIn}{\faGithub}. We then analyze how the Dark Sink modifies this benchmark, solving coupled Boltzmann equations for the dark-sector energy density and DM yield. We check the contribution of the Dark Sink $\psi$'s to dark radiation; consistency with existing data limits the maximum attainable cross section. For DM with a mass between $\text{MeV} -\mathcal{O}(10\text{ GeV})$, adding the Dark Sink can increase predictions for the direct detection cross section
all the way up to the current limits.
\end{abstract}

\maketitle
\preprint{FTPI-MINN-23-24}
\preprint{LCTP-23-17}

On-going direct detection experiments~\cite{XENON:2018voc,XENON:2019gfn,SENSEI:2023zdf,SENSEI:2020dpa,DarkSide:2022knj,PandaX:2022xqx,DAMIC-M:2023gxo,Arnquist:2023wmy} and a growing number of future proposals (see \emph{e.g.}~\cite{Essig:2022dfa} for an overview) promise greatly increased sensitivity over an expanding range of DM masses. A target of these experiments is the freeze-in benchmark~\cite{Hall:2009bx,Chu:2011be,Essig:2011nj} where DM is produced through a light dark photon mediator.  

The dark photon benchmark is well-motivated. A dark photon enjoys a privileged decoupling of constraints in the limit that it becomes massless~\cite{An:2013yfc}. Other light mediators that couple to electrons experience relatively tight constraints from stellar bounds~\cite{Hardy:2016kme}. Furthermore, for fermionic DM lighter than $\mathcal{O}(5 \text{ MeV})$, the successful predictions of Big Bang Nucleosynthesis (BBN) can present an obstacle to the construction of models with large direct detection cross sections \cite{Sabti:2019mhn}.\footnote{Models of such light DM which evade the BBN bounds include HYPERs~\cite{Elor:2021swj} and UV freeze-in at low reheating temperatures~\cite{Bhattiprolu:2022sdd}.} Given these arguments, a relevant question is: are the experiments targeting DM frozen-in via the dark photon probing other reasonable models of DM? Or are these direct detection experiments utilizing electron recoils largely testing a single idea? The substantial experimental effort motivates a concurrent effort by theorists to elucidate which models of DM are coming under the microscope.

In this Letter, we introduce a simple dark sector which modifies the predicted signals of non-thermal DM scenarios: \emph{the Dark Sink}. A Dark Sink transports energy from the DM into light dark sector states that do not contribute to the DM density. The representative Dark Sink we present is comprised of a neutral fermion $\psi$ which solely interacts with DM $\chi$. These interactions help determine the correct DM abundance via $\xxtopsps$ annihilations. 

We add this Dark Sink to the minimal freeze-in benchmark in which $\chi$ is charged under a gauged $U(1)'$ whose dark photon $\gamp$ kinetically mixes with the SM photon. As a byproduct of our analysis, we note the current literature for this freeze-in scenario contains errors~\cite{Chu:2011be,Essig:2011nj}.\footnote{There are papers which agree with our updated results and have noted some of the above errors, but do not give an easily accessible correction to the usual benchmark found widely in the literature~\cite{Fernandez:2021iti,Heeba:2023bik}.} We provide our corrected prediction for this model which is of immediate relevance as a primary target for ongoing direct detection experiments~\cite{XENON:2018voc,XENON:2019gfn,SENSEI:2023zdf,SENSEI:2020dpa,DarkSide:2022knj,PandaX:2022xqx,DAMIC-M:2023gxo,Arnquist:2023wmy}.

We detail the coupled Boltzmann equations of the Dark Sink and solve them numerically for the dark-sector energy density and DM yield. We find the range of possible direct detection cross sections for DM in the $\text{MeV} - \text{ TeV}$ mass range while ensuring the correct DM relic abundance and that $\psi$'s do not contribute too much to the effective number of cosmological neutrinos $\Neff$. For DM in the MeV to 100 GeV range, the power of the Dark Sink is to essentially allow any direct detection cross section between current experimental bounds and the freeze-in benchmark. Thus, just the existence of this extra state $\psi$ in one of the simplest models of DM can have significant consequences; any improvement in experimental bounds probes Dark Sink models. 
 
\mysec{The Dark Sink}To the SM, we add a gauged $U(1)'$ with dark fine structure constant $\alpha'$. The associated light dark photon $\gamp$ kinetically mixes with SM hypercharge. After electroweak symmetry breaking, the kinetic mixing to the SM photon is
\al{
\mathcal{L}\supset \frac{\epsilon}{2} F'_{\mu \nu} F^{\mu \nu}.
}
We also assume the dark photon has a negligible mass. For concreteness, we set $m_\gamp \sim 10^{-24} \text{ GeV}$ so that the kinetic mixing is unconstrained by COBE/FIRAS \cite{Fixsen:1996nj,Caputo:2020bdy} or black hole superradiance \cite{Baryakhtar:2017ngi,Siemonsen:2022ivj}. The lightness of the dark photon with respect to the energy transfers in direct detection experiments  enhances the direct detection cross section.  We consider Dirac fermionic DM $\chi$ with charge $+1$ under $U(1)'$ in the range $\text{MeV} \lesssim m_\chi \lesssim \text{TeV}$. The lower bound is motivated by the threshold of ongoing direct detection experiments, but also allows us to ignore plasmon decay contributions to freeze-in \cite{Dvorkin:2019zdi}. The upper bound is set by perturbativity considerations, as we discuss in depth below. As in the usual case of freeze-in, it is helpful to define the portal coupling, $\kappa \equiv \epsilon \sqrt{\alpha'/\alpha}$, which determines both the amount of DM production from SM thermal bath annihilations and the expected direct detection rates. 

The Dark Sink augments the freeze-in benchmark through the introduction of a light, neutral dark fermion $\psi$. $\psi$ interacts with $\chi$ but, importantly, not with anything in the SM. The $\psi-\chi$ interaction is mediated by a heavy scalar mediator $\Phi$:
\al{
\mathcal{L} \supset \frac{y_\chi y_\psi}{m_\Phi^2} \overline{\chi} \chi \overline{\psi} \psi.
\label{eq:chipsicoupl}
}
We assume $\Phi$ is sufficiently heavy so that it is produced negligibly and the effective operator in Eq.~\eqref{eq:chipsicoupl} is sufficient. This is an assumption that can be readily satisfied for sub-GeV DM where $m_\Phi \gtrsim 50 m_\chi$ still allows both the correct relic abundance and perturbative couplings. However, making DM annihilations large enough for heavier DM starts to require lighter $\Phi$, a point we return to later. We have also verified that the DM self interactions mediated by $\Phi$ are sufficiently small.

We must also ensure that $\psi$ is not too heavy or abundant so that it does not contribute substantially to the DM relic abundance or $\Neff$.\footnote{A simpler model where $\chi$'s annihilate to light scalar $\phi$'s via a Yukawa interaction---without the addition of the fermionic $\psi$---fails because the Yukawa interaction also gives rise to a too large $\chi$ self-interaction for sub-GeV DM.} For the former, it is sufficient and simplest to assume $\psi$ has a negligible mass, as we do for the rest of the paper.  The latter gives a constraint on the parameter space for the Dark Sink, which we will explicitly verify.

The coupling in Eq.~\eqref{eq:chipsicoupl} allows $\xxtopsps$. The DM production proceeds in 2 simultaneous steps: 1) $\smsmtoxx$ annihilations of charged SM particles produce DM pairs through the vector portal. 2) $\chi$'s quickly thermalize with $\psi$'s to a dark-sector temperature $T' < T$, eventually annihilating via $\xxtopsps$ to deplete the DM abundance until it reaches the observed value.

\mysec{Boltzmann Equations and Solutions}We begin by enumerating the set of coupled Boltzmann equations which govern the evolution of the energy density in the dark sector and the DM yield, assuming Maxwell-Boltzmann statistics. First, the energy density in the dark sector, due to $\smsmtoxx$ processes through the vector portal, is governed by the Boltzmann equation:
\begin{widetext}
\al{
-\overline{H} T \frac{d \rho'}{d T} + 3 H (\rho' + p')
&= 
\sum_{(i, j)}
\frac{4 g_i^2}{(4 \pi)^5} \int_{s_\text{min}}^{\infty} \!  ds \ \overline{\left|\mathcal{M}\right|}^2_{i j \rightarrow \chi \overline{\chi}}
\sqrt{s - 4 m_i^2} \sqrt{s - 4 m_\chi^2}
\left(
T K_2 \left(\frac{\sqrt{s}}{T}\right)
-
T' K_2 \left(\frac{\sqrt{s}}{T'}\right)
\right),\notag\\
\quad \text{with} \quad
H/\overline{H} &= 1 + \frac{1}{3} \frac{d \ln g_{\ast, s}}{d \ln T} + \frac{1}{3} \frac{d \ln g_{\ast, s}}{d \ln T'} \frac{T}{T'} \frac{d T'}{d T} \quad \text{and} \quad
p' =  \frac{\rho'}{3} - \frac{m_\chi^3 T'}{3 \pi^2} K_1\left(\frac{m_\chi}{T'}\right),
\label{eq:drhopdT}
}
\end{widetext}
where we sum over $(i, j) = (f, \overline{f})$, $(\pi^+, \pi^-)$, $(K^+, K^-)$, $(W^+, W^-)$. $g_i$ is the number of degrees of freedom for the SM particle $i$, $\overline{\left|\mathcal{M}\right|}^2$ is the fully-averaged squared matrix element integrated over $\cos \theta$ in the center-of-mass frame, and $s_\text{min} = \text{max}(4 m_i^2, 4 m_\chi^2)$. We have assumed Maxwell-Boltzmann statistics in deriving $p'$ above.

\begin{figure}[t!]
\centering
\includegraphics[width=0.98 \columnwidth]{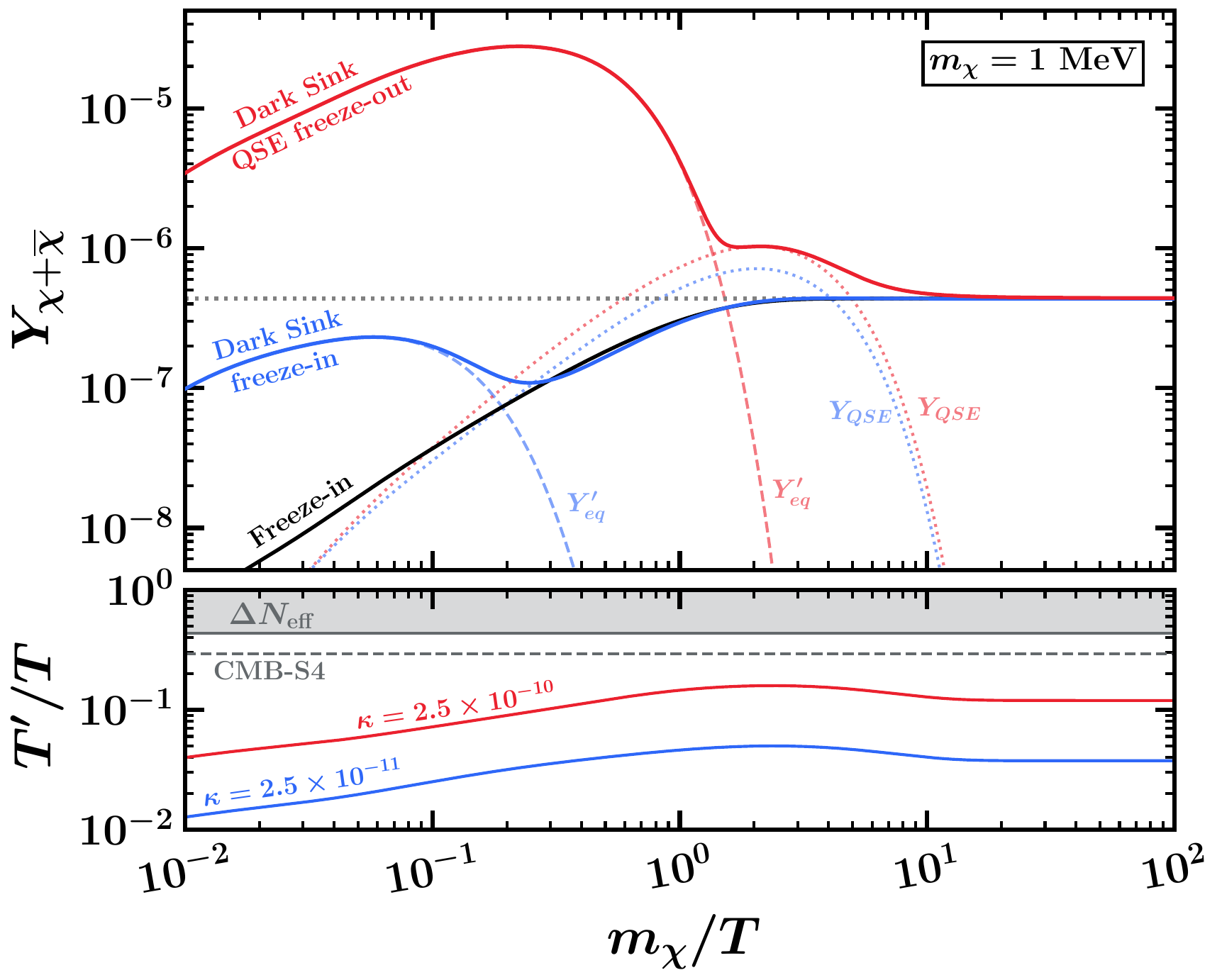}
\caption{The evolution of the DM yield (top panel) and the temperature in the dark sector relative to the SM bath (bottom panel), for $m_\chi = 1$ MeV and $\kappa = 2.5 \times 10^{-10}$ ({\color[HTML]{EB212E} red}) and $2.5 \times 10^{-11}$ ({\color[HTML]{2E67F8} blue}), as a function of $m_\chi/T$. The red and blue curves correspond to two qualitatively different ways to achieve the observed DM relic abundance (dotted gray): ``Dark Sink QSE freeze-out" and ``Dark Sink freeze-in", respectively. The usual freeze-in curve ($ \kappa_\text{FI} = 1.94 \times 10^{-11}$; black; top panel) is also shown for comparison. See text for more details.}
\label{fig:YieldMeV}
\end{figure}

The evolution of the DM number density, or equivalently DM yield $Y \equiv n_\chi/s$ defined as the ratio of the DM number density $n_\chi$ and the total entropy in the visible and the dark sectors $s$, is governed by
\al{
-\frac{\overline{H} T}{s} \frac{d Y}{d T} \! = \! \sigv' \left[ {Y'_{eq}}^2 + \left(1 - \frac{Y^2}{Y_{eq}^2}\right) Y_{QSE}^2 - Y^2 \right].
\label{eq:dYovrdT}
}
$Y_{QSE} = Y_{eq} \sqrt{\frac{\sigv}{\sigv'}}$ represents a quasi-static equilibrium abundance~\cite{Cheung:2010gj,Chu:2011be} described in more detail below, and the thermally-averaged annihilation cross section for $\xxtopsps$ is
\begin{equation}
\label{eq:sigmaPsiChi}
\sigv' \approx \frac{3}{4 \pi} \frac{y_\chi^2 y_\psi^2}{m_\Phi^4} m_\chi T'.
\end{equation}
$\sigv$ is the thermally-averaged annihilation cross section of DM to the SM summed over all final-state pairs of charged SM fermions. Here we have taken the limit $m_\chi/T' \gg 1$, valid for all later times of interest during the DM's evolution. 

\begin{figure}[t!]
\centering
\includegraphics[width=0.98 \columnwidth]{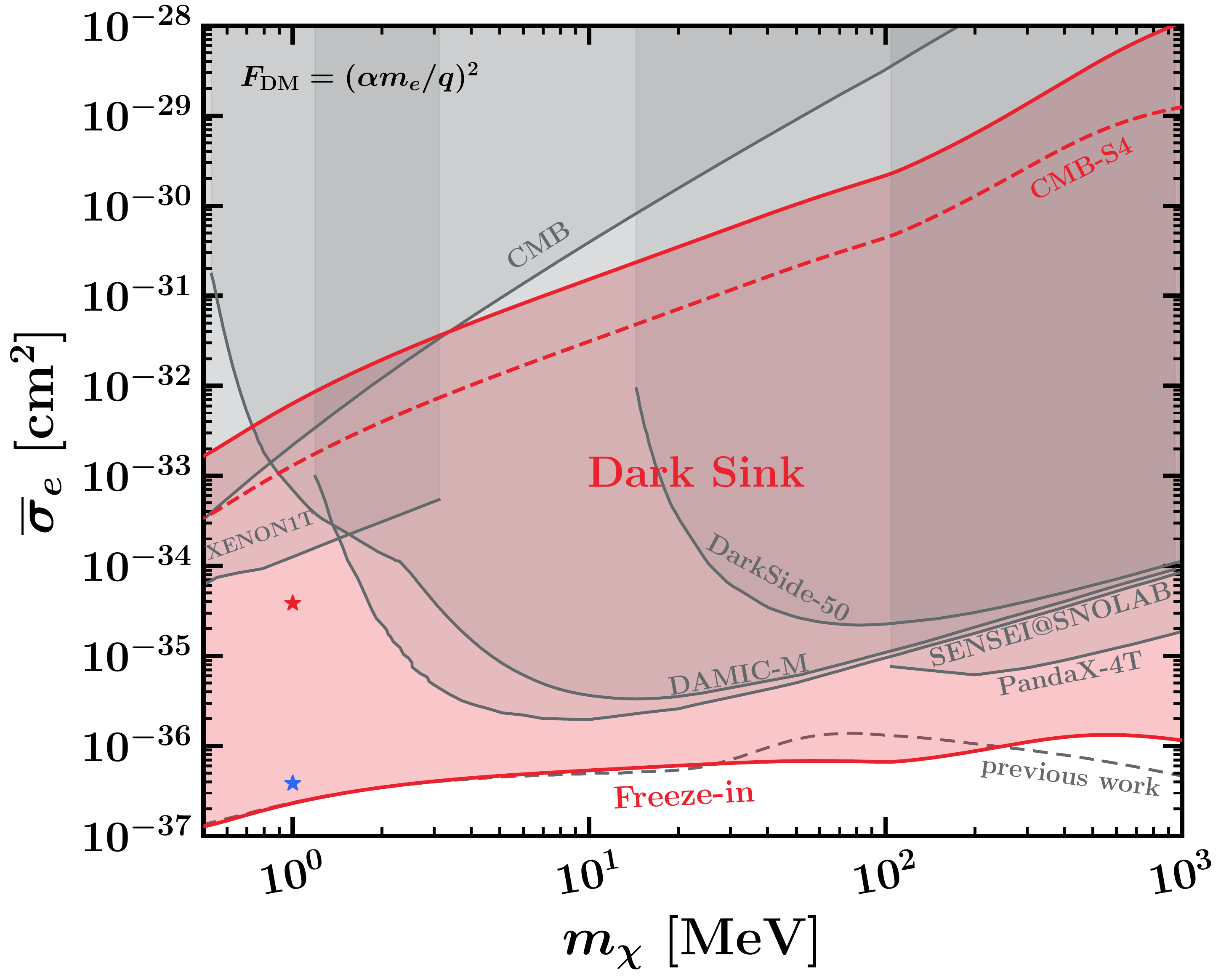}
\caption{The power of the Dark Sink is to lift the usual freeze-in benchmark such that the entire {\color[HTML]{EB212E} red} region reproduces the correct relic abundance. Shown in grey are the latest direct detection constraints from PandaX~\cite{PandaX:2022xqx}, DAMIC-M~\cite{DAMIC-M:2023gxo,Arnquist:2023wmy}, SENSEI~\cite{SENSEI:2023zdf,SENSEI:2020dpa}, XENON1T S2 data \cite{XENON:2019gfn} from solar reflected DM \cite{An:2021qdl}, and DarkSide~\cite{DarkSide:2022knj}, as well as constraints from the CMB~\cite{Buen-Abad:2021mvc,Planck:2018vyg,CMB-S4:2016ple}. The previous result for the freeze-in benchmark is shown in dashed grey \cite{Essig:2011nj}.}
\label{fig:SubGeVDarkSink}
\end{figure}

\begin{figure}[t!]
\centering
\includegraphics[width=0.98 \columnwidth]{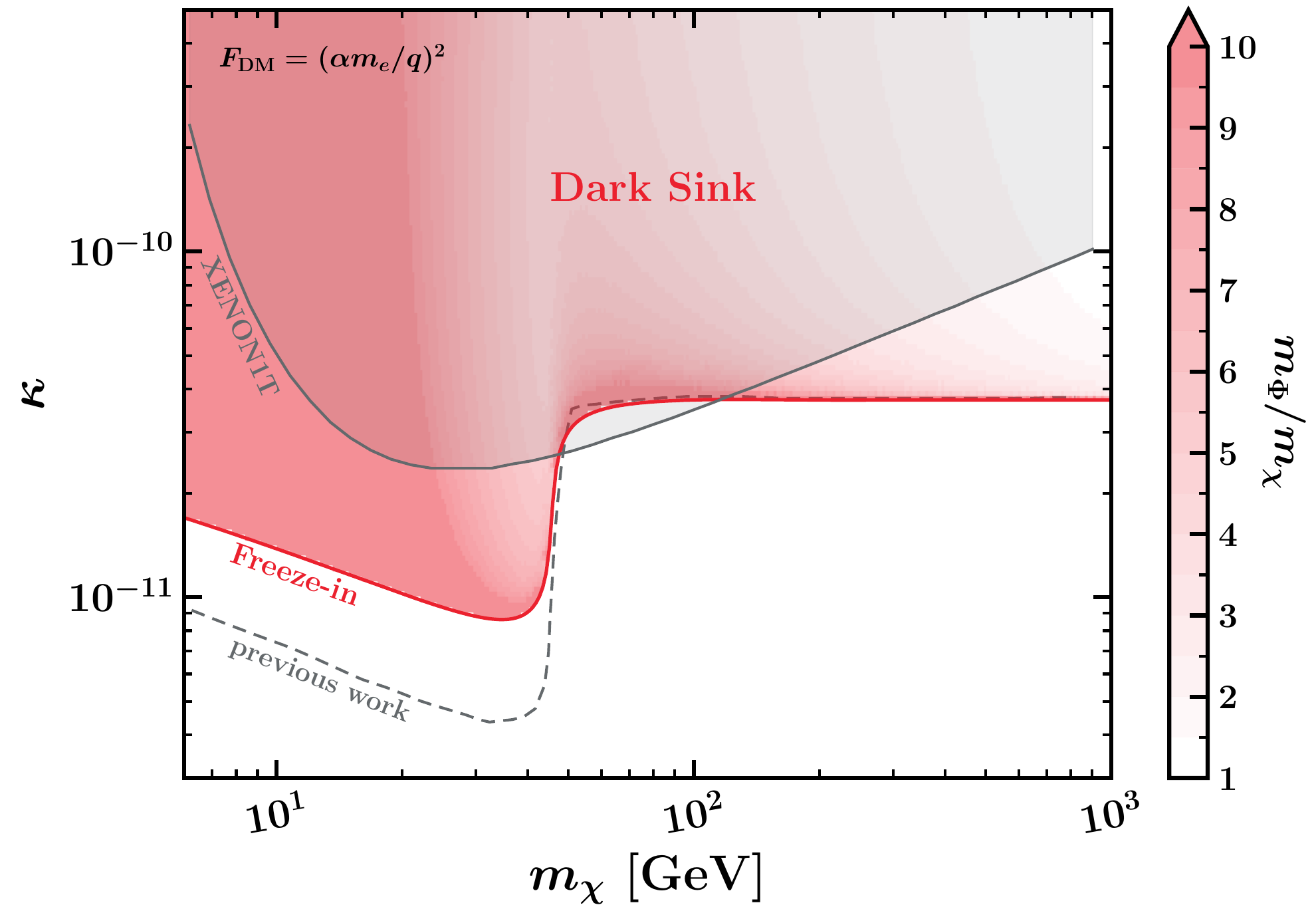}
\caption{The power of the Dark Sink is to lift the usual freeze-in benchmark such that the entire {\color[HTML]{EB212E} red} region reproduces the correct relic abundance. Shown in grey is the latest direct detection constraint from XENON1T~\cite{XENON:2018voc,Hambye:2018dpi} as well as the perturbativity constraint discussed in the text. The previous result for the freeze-in benchmark is shown in dashed grey \cite{Chu:2011be}.}
\label{fig:GeVDarkSink}
\end{figure}

Having enumerated the Boltzmann equations for the Dark Sink scenario, let us take the limit in which $y_\psi \to 0$ so that the Dark Sink decouples from DM. Doing so, the number density Boltzmann equation simplifies to
\al{
-\frac{\overline{H} T}{s} \frac{d Y}{d T} = \sigv {Y_{eq}}^2.
\label{eq:FIdYdT}
}
This recovers the usual freeze-in scenario and the resulting prediction is shown in black in Fig.~\ref{fig:YieldMeV} and as the (bottom) solid red lines in Figs.~\ref{fig:SubGeVDarkSink} and \ref{fig:GeVDarkSink}. This differs from the prediction often cited for this scenario~\cite{Chu:2011be,Essig:2011nj}, shown as dashed gray curves. The discrepancy for $m_\chi > 1 \text{ GeV}$ may be traced to an incorrect factor in going from the gauge to mass basis for the (dark) photons, while the source of the discrepancy for $m_\chi < 1 \text{ GeV}$ is still unknown. For details, see the Supplementary Material.

We now turn to the impact of the Dark Sink. It accommodates larger values of $\kappa$, which \emph{a priori} would over-produce the DM; the $\chi-\psi$ interaction provides compensating annihilations and we adjust $\sigv'$ to ensure the correct DM abundance is recovered. Depending on how much greater $\kappa$ is than the freeze-in value, $\kappa_{FI}$, different cosmological histories follow. For much larger $\kappa$, we dub the qualitative behavior ``Dark Sink QSE freeze-out.'' An example is shown as the red curve in Fig.~\ref{fig:YieldMeV} and may be understood as follows. At early times, $Y_{QSE} \ll Y'_{eq}$ since $\sigv \ll \sigv'$. During this period, the middle term in Eq.~\eqref{eq:dYovrdT} is negligible relative to the first term. Omitting it, we find the Boltzmann equation resembles that of ordinary freeze-out. Thus, the DM yield traces $Y'_{eq}$ until it begins to freeze-out. At this point, we transition to a regime where $Y_{QSE} \gtrsim Y'_{eq}$. So, we may instead ignore the first term in Eq.~\eqref{eq:dYovrdT} for the middle term. The Boltzmann equation again resembles that of ordinary freeze-out, but now in which $Y_{QSE}$ plays the role of the usual equilibrium yield. Then, the DM yield traces $Y_{QSE}$ until the annihilations $\chi \overline{\chi} \to \psi \overline{\psi}$ become slow relative to the Hubble expansion rate. This occurs roughly when $\sigv' Y \sim H/s$.  This second freeze-out can then result in the observed DM abundance. 

The above occurs as long as there is a sufficient buildup of DM to allow it to follow $Y_{QSE}$ during the intermediate range of temperatures. However, if $\kappa$ is relatively close to $\kappa_{FI}$, then the evolution is qualitatively different. In this case, DM undergoes what we call ``Dark Sink freeze-in'', an example of which is shown as the solid blue curve in Fig.~\ref{fig:YieldMeV}.  Here, at early times, again $Y_{QSE} \ll Y'_{eq}$, the middle term in Eq.~\eqref{eq:dYovrdT} is negligible, and the DM yield traces $Y'_{eq}$ until it begins to freeze-out. However, if there is not enough DM at this freeze-out time, then annihilations of SM particles to DM pairs are not balanced by DM annihilations to $\psi$ pairs. Then, both the first and last terms on the right side of Eq.~\eqref{eq:dYovrdT} are negligible, and the middle term may be rewritten as simply $\sigv {Y_{eq}}^2$. This corresponds to the usual Boltzmann equation for freeze-in. The only difference to ordinary freeze-in is: the initial epoch where $Y$ traced  $Y_{eq}'$ causes the initial DM yield to be slightly smaller than the would be yield of a pure freeze-in scenario at the same $T/m_{\chi}$. This indicates that $\kappa$ must be slightly larger than in the usual freeze-in paradigm in order to achieve the correct relic abundance. See the Supplementary Material for the evolution of the yield for benchmarks with different DM masses.

The joint Planck CMB and baryon acoustic oscillation (BAO) measurements of $N_\text{eff}$ constrain the dark-sector temperature to be \cite{Planck:2018vyg,Cielo:2023bqp}
\al{
T^\prime/T < 0.437 \quad \text{(95\% CL)},
\label{eq:Tpbnd}
}
while upcoming CMB-S4 experiment is expected to further constrain $T^\prime/T$ to less than 0.292 at 95\% CL \cite{CMB-S4:2016ple}. In Fig.~\ref{fig:YieldMeV}, we show the (expected) $\Neff$ bound on $T^\prime/T$ from Planck (CMB-S4) as a solid (dashed) gray line in the bottom subplot. There, we also show the evolution of $T^\prime/T$ as a function of $m_\chi/T$, after numerically integrating Eq.~\eqref{eq:drhopdT}, for two benchmark $\kappa$.

\mysec{The power of the Dark Sink}
The existence of the light $\psi$ may dramatically impact the expected $\chi$ direct detection signals as the $\psi$ annihilation channel largely decouples the DM relic abundance from the expected direct detection rate. The rate of direct detection can greatly exceed the usual freeze-in expectation. 

The direct detection cross section through the light $\gamp$ mediator at the usual reference momentum is \cite{Essig:2015cda}
\al{
\sige = \frac{16 \pi \mu_{\chi e}^2 \alpha^2 \kappa^2}{\prn{\alpha m_e}^4}.
}
The range of direct detection cross sections allowed by the Dark Sink scenario are shaded red in Figs.~\ref{fig:SubGeVDarkSink} and \ref{fig:GeVDarkSink}. In this region, the coupling of $\chi$'s to $\psi$'s is perturbative and gives the correct relic abundance. We also ensure compliance with $\Neff$ as follows. We find the largest $\kappa$ for a given $m_\chi$ which satisfies Eq.~\eqref{eq:Tpbnd} by numerically integrating Eq.~\eqref{eq:drhopdT} and deduce the resulting upper bound on $\overline{\sigma}_e$. The top solid (dashed) red line in Fig.~\ref{fig:SubGeVDarkSink} corresponds to the (expected) 95\% CL upper limit on $\overline{\sigma}_e$ from the $N_\text{eff}$ measurements by Planck (CMB-S4). Also shown in gray are the current direct detection constraints from PandaX~\cite{PandaX:2022xqx}, DAMIC-M~\cite{DAMIC-M:2023gxo,Arnquist:2023wmy}, SENSEI~\cite{SENSEI:2023zdf,SENSEI:2020dpa}, DarkSide~\cite{DarkSide:2022knj}, and XENON1T~\cite{XENON:2019gfn,An:2021qdl,XENON:2018voc,Hambye:2018dpi}. Constraints from CMB+BAO~\cite{Buen-Abad:2021mvc} become competitive with these for $m_\chi \lesssim \text{ MeV}$. These constraints consider how DM-SM interactions cool baryons and exert pressure on DM. The resulting earlier recombination and suppressed structure formation modifies the CMB spectra and matter power spectrum. 

In Fig.~\ref{fig:SubGeVDarkSink}, we see that the Dark Sink can allow cross sections in a region of parameter space which is being actively probed, but for which there are few other known models due to stringent cosmological and astrophysical constraints. For illustration, the chosen values of $\kappa$ corresponding to Dark Sink QSE freeze-out and Dark Sink freeze-in shown in Fig.~\ref{fig:YieldMeV} are denoted by stars. 

In Fig.~\ref{fig:GeVDarkSink}, again we see that the Dark Sink is being actively probed by ongoing direct detection and gives further interesting benchmarks between the freeze-in line and current bounds. Notably, there is no top boundary corresponding to an $\Neff$ constraint as in the sub-GeV case. For these heavier DM, $\psi$'s are always decoupled and redshifting as radiation before the QCD phase transition, which then guarantees their contribution to $\Neff$ is below current constraints. Another difference to the light DM case is shown as a gradient for $m_\chi \gtrsim 20 \text{ GeV}$. As in the WIMP paradigm, for heavier masses, achieving a sufficiently large cross section $\sigv'$ begins to require non-perturbative couplings. To offset larger couplings, we are pressed to consider lighter $\Phi$. As $m_\Phi$ approaches $m_\chi$, some $\Phi$ would be produced on shell by DM scatters in the dark sector bath. Then, a more proper analysis tracing the $\Phi$ abundance and contribution to the dark bath is needed.

Though a more thorough treatment is necessary for these heavier $m_\chi$, it does not present any insurmountable challenges. One would need to dynamically track the yield of $\Phi$, analogous to Eq.~\eqref{eq:dYovrdT}. Since $\Phi$ only couples to $\psi$'s and $\chi$'s via Yukawa couplings, the most relevant processes on the right side of such an equation would be decays and inverse decays to pairs of these fermions. For $T' \gtrsim m_\Phi$, $\Phi$'s would have a non-negligible abundance in the dark-sector bath and would contribute to $g'_{\ast,s}$. As $T'$ drops below $m_\Phi$, the Yukawas $y_\chi$ and $y_\psi$ would determine the relative branching ratios of $\Phi$'s. If there is sufficient time before $T'$ drops below $m_\chi$, this relative branching will be erased by the thermalization of DM with the Dark Sink. However, if $m_\Phi$ is too close to $m_\chi$,  thermalization may be incomplete and the decays of $\Phi$ may leave an imprint on the evolution of the DM abundance.

While constraints coming from large-scale coherent magnetic fields and plasma instabilities have significant uncertainties at present, in the future, \cite{Stebbins:2019xjr,Lasenby:2020rlf} and related approaches may rule out the possibility of $\chi$ comprising all of DM. Should $\chi$ only make up a sub-component of DM, $\kappa_\text{FI}^\text{sub}$ would be proportionately smaller resulting in an even larger parameter space than the range shown in Fig.~\ref{fig:SubGeVDarkSink}.

\mysec{Discussion}In this Letter, we have introduced the Dark Sink: light degrees of freedom in the dark sector exclusively coupled to DM and not to the mediator or the SM itself. For simplicity, we have taken a single light dark fermion $\psi$ to fill this role and demonstrated that the Dark Sink elevates the difficult-to-detect phenomenology of usual freeze-in benchmarks to detectable heights. The power of the Dark Sink is highlighted in Figs.~\ref{fig:SubGeVDarkSink} and \ref{fig:GeVDarkSink} where current direct detection experiments are found to be probing Dark Sink parameter space. 

We have focused on DM masses in the $\text{MeV} - \text{TeV}$ range. The upper bound preserves perturbativity of our Dark Sink, while the lower one is more arbitrary. Below $m_\chi \sim \text{ MeV}$, plasmon decay contributions to freeze-in become important \cite{Dvorkin:2019zdi}. Accounting for this process in Eq.~\eqref{eq:drhopdT} should be straightforward and could yield Dark Sink models at even lower masses. Although we have concentrated on ongoing direct detection efforts, these lower-mass benchmarks would be relevant for a host of proposed experiments targeting such sub-MeV DM \cite{Hochberg:2019cyy,Geilhufe:2019ndy,Griffin:2020lgd,Hochberg:2021pkt,Knapen:2021run,Hochberg:2021ymx}; we leave this to future work.

While we have illustrated the impact of a single additional light dark sector particle in a  well-motivated example with implications for direct detection, it also of interest to study a Dark Sink's impact on other freeze-in phenomenology, for example, long-lived particle~\cite{Co:2015pka} searches. A Dark Sink could also modify direct detection signals in UV freeze-in scenarios~\cite{Elor:2021swj,Bhattiprolu:2022sdd}, though care may be required~\cite{Forestell:2018dnu}. We leave these directions for future work.

\begin{acknowledgments}
We thank Rouven Essig, David E.~Morrissey, Hitoshi Murayama, Katelin Schutz, and Jessie Shelton for useful discussions. This work is supported in part by the DoE grant DE-SC0007859. RM thanks the Mainz Institute for Theoretical Physics (MITP) of the Cluster of Excellence PRISMA${}^+$ (project ID 39083149) for their hospitality while a portion of this work was completed.

\end{acknowledgments}
\bibliography{ref}

\clearpage

\onecolumngrid
\begin{center}
  \textbf{\large Supplementary Material for A Dark Sink Enhances the Direct Detection of Freeze-in Dark Matter}
\\[.2cm]
  \vspace{0.05in}
  {Prudhvi N.~Bhattiprolu, \ Robert McGehee, \ and \ Aaron Pierce}
\end{center}

\twocolumngrid
\setcounter{equation}{0}
\setcounter{figure}{0}
\setcounter{page}{1}
\makeatletter
\renewcommand{\theequation}{S\arabic{equation}}
\renewcommand{\thefigure}{S\arabic{figure}}

\section{Vector Portal Freeze-In}

To the SM Lagrangian, we add a kinetically mixed (and effectively massless) dark photon as well as DM
\al{
\mathcal{L} \supset - \frac{1}{4} \hat{X}_{\mu \nu} \hat{X}^{\mu \nu} + \frac{\epsilon_Y}{2} \hat{X}_{\mu \nu} \hat{B}^{\mu \nu} - e' \hat{X}_\mu \overline{\chi} \gamma^\mu \chi.
}
For simplicity, we have normalized the charge of DM under $A'_\mu$ to $1$. After diagonalization, the gauge-basis vectors $\{\hat{A}_\mu,\hat{Z}_\mu,\hat{X}_\mu\}$ written in terms of the mass basis $\{A_\mu,Z_\mu,A'_\mu\}$ are
\al{
\hat{Z}_\mu &= Z_\mu \nonumber \\
\hat{A}_\mu &= A_\mu + \epsilon A'_\mu \\
\hat{X}_\mu &= A'_\mu - \epsilon \tan \theta_W Z_\mu. \nonumber
}
In the above, we have taken the small $\epsilon$ limit and expressed mixing in terms of the effective mixing between the SM and dark photons, $\epsilon \equiv \epsilon_Y \cos\theta_W$. The $\tan \theta_W$ in the final equation is the  source of our discrepancy with the previous result for the freeze-in benchmark for heavier DM \cite{Chu:2011be}, which omitted this factor in Eq.~(73) of \cite{Chu:2011be}, as shown in Fig.~3. This error was previously pointed out in \cite{Heeba:2023bik}.
The interaction terms expressed in this mass basis are
\al{
\mathcal{L} \supset &- \epsilon e A'_\mu J^\mu_\text{EM} - e' J^\mu_\text{DM} \prn{A'_\mu - \epsilon \tan \theta_W Z_\mu} \\
&+i \epsilon e \left[ F^{\prime \mu \nu} W_\mu^+ W_\nu^- - \prn{\partial_\mu W_\nu^+ - \partial_\nu W_\mu^+}A^{\prime \mu} W^{- \nu} \right. \nonumber \\
&+ \left. \prn{\partial_\mu W_\nu^- - \partial_\nu W_\mu^-}A^{\prime \mu} W^{+ \nu} \right]. \nonumber
}
The SM electromagnetic (EM) current picks up a ``milli-charge'' under $U(1)'$ while the dark current picks up an analogous small charge to the SM $Z$. A different rotation choice is possible in the massless dark photon limit which instead causes the dark current to pick up a ``milli-charge'' under the SM $U(1)_\text{EM}$~\cite{Fabbrichesi:2020wbt}. While either choice is valid, the one we have used also corresponds to the couplings in the case of a massive dark photon. The SM $W$ bosons also inherit a coupling to the dark photon. 

Freezing-in DM through the vector portal proceeds via $f \overline{f}$ annihilations of SM fermion pairs, $\phi^+ \phi^-$ annihilations of SM scalar pairs (such as pions below $\Lambda_\text{QCD}$), and $W^+ W^-$ annihilations. While the former two sets of processes are commonplace in the literature, the latter are usually not included due to their sub-leading contributions.
For completeness, we include the fully-averaged squared matrix elements for all these processes:
\begin{widetext}
\al{
\begin{split}
\overline{\left| \mathcal{M}\right|}^2_{f \overline{f} \to \chi \overline{\chi}}
 ={}& 
\frac{32}{3} \pi^2 \alpha^2 \kappa^2 N_f \prn{s + 2 m_\chi^2}
\left[
\vphantom{\frac{V_f^2 \prn{s + 2 m_f^2} + A_f^2 \prn{s - 4 m_f^2}}{\prn{s - m_Z^2}^2 + m_Z^2 \Gamma_Z^2}}
\right.
\frac{Q_f^2}{s^2} \prn{s + 2 m_f^2}
-
2 Q_f V_f \tan \theta_W \frac{\prn{s + 2 m_f^2} \prn{s - m_Z^2}}{s\left[\prn{s - m_Z^2}^2 + m_Z^2 \Gamma_Z^2\right]}
\\&
\hphantom{\frac{32}{3} \pi^2 \alpha^2 \kappa^2 N_f \prn{s + 2 m_\chi^2} [}
+
\left.
\tan^2\theta_W \frac{V_f^2 \prn{s + 2 m_f^2} + A_f^2 \prn{s - 4 m_f^2}}{\prn{s - m_Z^2}^2 + m_Z^2 \Gamma_Z^2}
\right],
\end{split}\\
\begin{split}
\overline{\left| \mathcal{M}\right|}^2_{\phi^+ \phi^- \to \chi \overline{\chi}}
 ={}& 
\frac{32}{3} \pi^2 \alpha^2 \kappa^2 \prn{1 + \frac{2 m_\chi^2}{s}} \prn{1 - \frac{4 m_\phi^2}{s}},
\end{split}\\
\begin{split}
\overline{\left| \mathcal{M}\right|}^2_{W^+ W^- \to \chi \overline{\chi}}
 ={}& 
\frac{8}{27} \pi^2 \alpha^2 \kappa^2 \prn{\frac{m_Z}{m_W}}^4 \frac{\prn{s+2 m_\chi^2} \prn{s-4 m_W^2} \prn{s^2+20 s m_W^2 +12 m_W^4}}{s^2 \left[ \prn{s-m_Z^2}^2 +m_Z^2 \Gamma_Z^2 \right]},
\end{split}
}
\end{widetext}
where $Q_f$ is the EM charge of $f$, 
$N_f =$ 1 for leptons or 3 for quarks, and
$V_f$ ($A_f$) is the vector (axial) coupling of the $Z$-boson to fermion pairs divided by the EM coupling $e$.
We have dropped a term proportional to $\Gamma_Z^2/m_Z^2$ for the $W^+ W^- \rightarrow \chi \overline{\chi}$ process.

\begin{figure*}[t!]
  \begin{minipage}[]{\columnwidth}
    \centering
    \includegraphics[width=0.98\columnwidth]{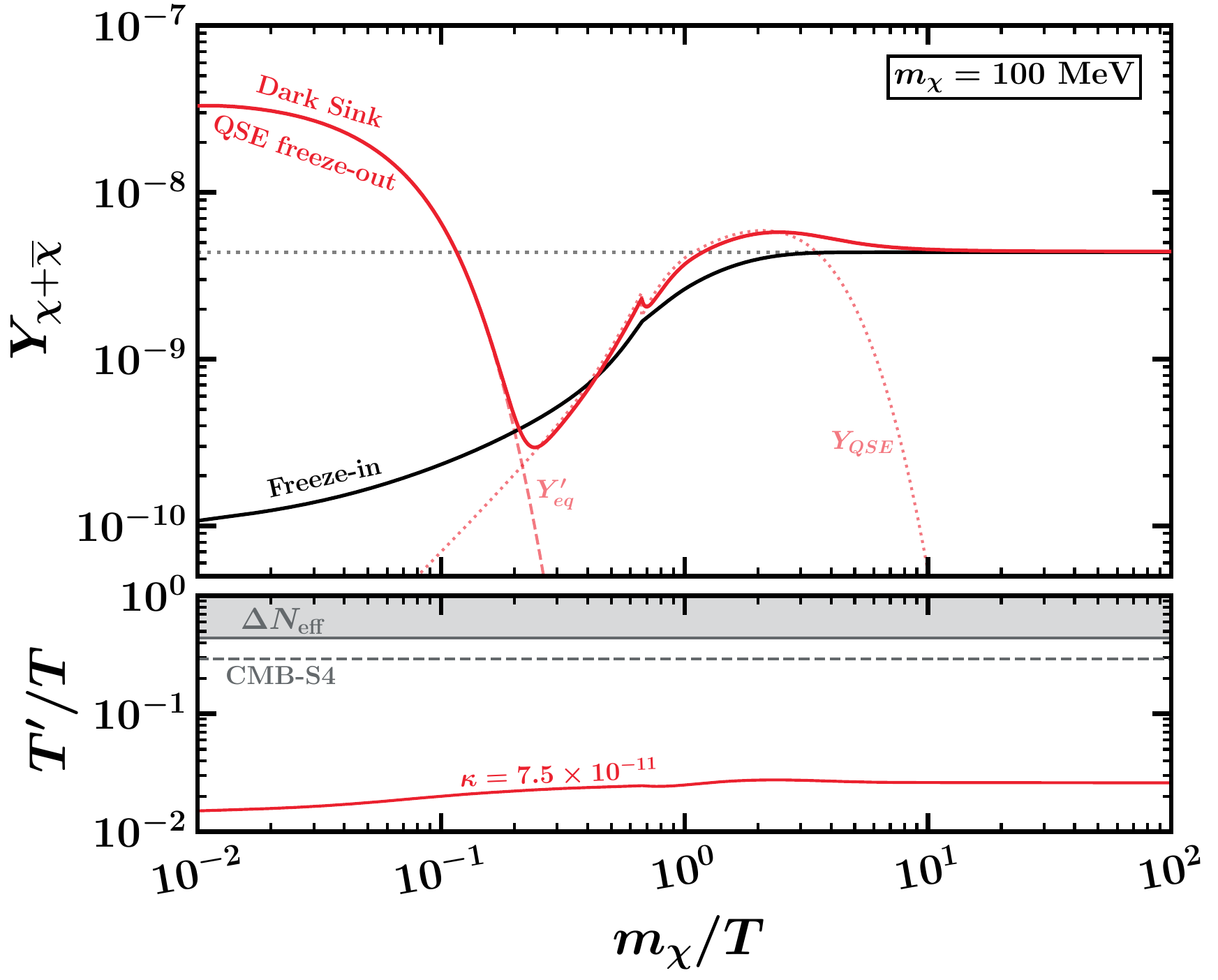}
  \end{minipage}
  \begin{minipage}[]{\columnwidth}
    \centering
    \includegraphics[width=0.98\columnwidth]{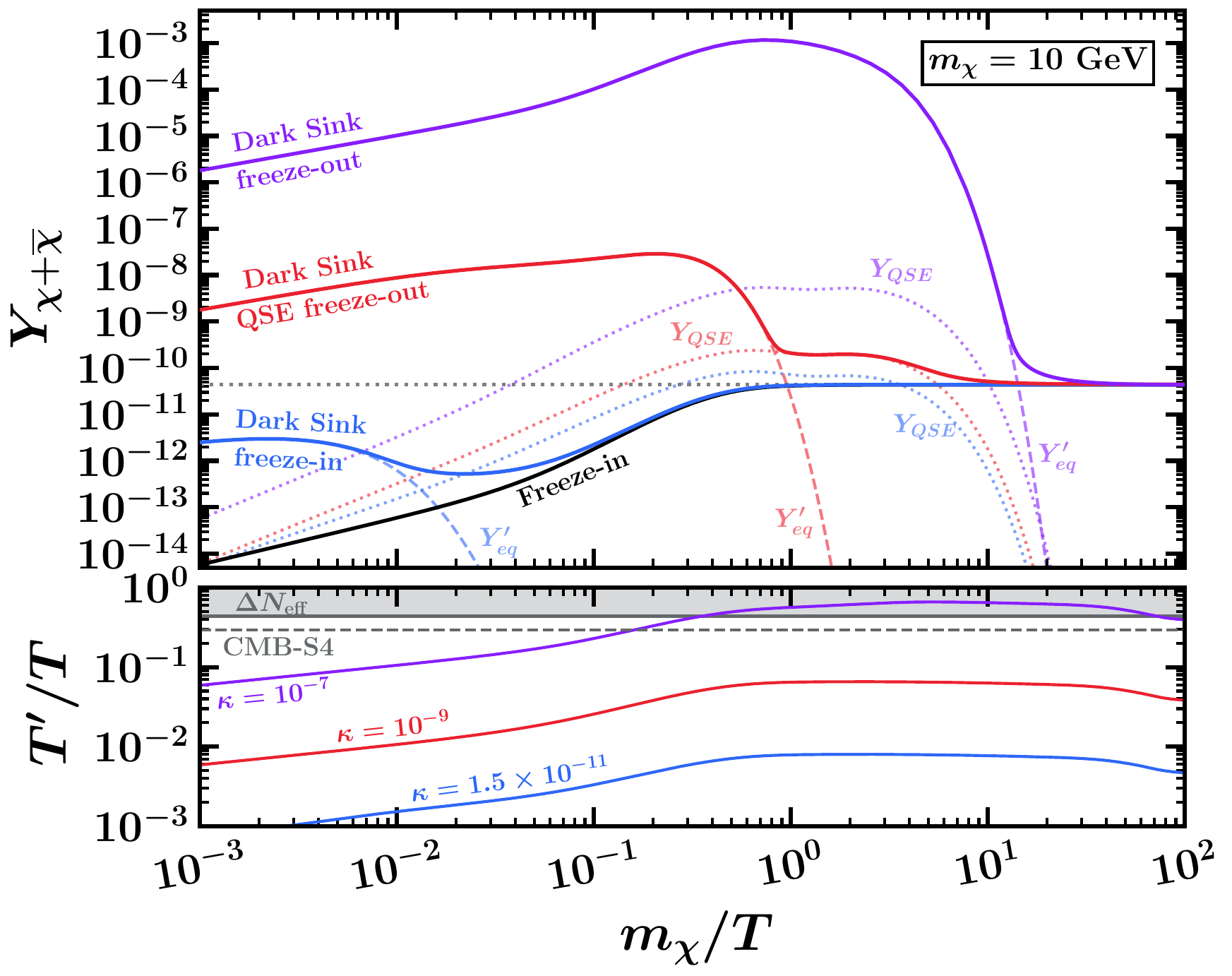}
  \end{minipage}
 \caption{Same as Fig.~1 for (Left) $m_\chi = 100 \text{ MeV}$ and (Right) $m_\chi = 10 \text{ GeV}$. For $m_\chi = 100 \text{ MeV}$, the chosen $\kappa$ illustrates a yield evolution which is at the boundary between Dark Sink freeze-in and Dark Sink QSE freeze-out. For $m_\chi = 10 \text{ GeV}$, we've also included a yield curve for such a large $\kappa$ that the evolution behaves just like freeze-out in the Dark Sink ({\color[HTML]{871efe} purple}). 
\label{fig:YieldSuppl}}
\end{figure*}

Since $\alpha'$ is small to avoid DM self interactions, a negligible amount of dark photons are produced and no significant amount of DM production (or annihilation) involves the dark photons. In the absence of the Dark Sink or additional interactions, the number-density Boltzmann equation for DM in Eq.~(4) simplifies to
\al{
-\frac{\overline{H} T}{s} \frac{d Y}{d T} = \sigv {Y_{eq}}^2,
}
where there is an implicit sum over the three sets of SM annihilations discussed above which freeze-in DM pairs. Numerically integrating and requiring the resulting yield reproduces the observed DM abundance yields the red freeze-in curves appearing in Figs.~2 and~3. 

As discussed above, there is a simple explanation for the discrepancy between our result and the existing literature for $m_\chi > 1 \text{ GeV}$. For the lighter masses, we are unable to reproduce the result from \cite{Essig:2011nj}. Variations in the parameterization of the QCD phase transition could not account for the discrepancy.

\section{Dark Sink freeze-in, QSE freeze-out, and freeze-out}

For illustrative purposes, we include the DM yield evolution for two more masses in Fig.~\ref{fig:YieldSuppl} in addition to the $m_\chi = 1 \text{ MeV}$ benchmark discussed in the main text and shown in Fig.~1. For $m_\chi = 100 \text{ MeV}$, we have chosen a $\kappa$ which results in a yield evolution at the boundary between Dark Sink freeze-in and Dark Sink QSE freeze-out. The noticeable kink near $m_\chi / T \sim 0.6$ corresponds to the QCD phase transition and its accompanying dilution of the relic abundance. For $m_\chi = 10 \text{ GeV}$, we have included the usual cases of Dark Sink freeze-in and Dark Sink QSE freeze-out discussed in the main text. We have also added a yield curve for a $\kappa$ large enough to cause an ordinary freeze-out evolution in the Dark Sink in purple for comparison.

As discussed in the main text, for the heavier $m_\chi = 10 \text{ GeV}$ DM, it is clear why there is no upper bound on $\kappa$ coming from $\Neff$. Even in the case of a large $\kappa = 10^{-7}$ which practically thermalizes the dark sector and SM, the current $\Neff$ constraints are avoided thanks to the large entropy dump occurring at the QCD phase transition. This occurs after the DM yield is set and $\psi$'s are just redshifting as radiation and causes the initially-too-large $\Neff$ to decrease to within the allowed range before BBN.

\end{document}